\documentclass{aastex631}
\usepackage{amsmath}
\usepackage{comment}
\usepackage{multirow}
\begin{document}

\title{High-resolution Observation of Mini-Filament Eruptions Near Coronal Hole Boundary and Their Response in Solar Corona}

\author[0000-0001-9049-0653]{Nengyi Huang}
\affiliation{Institute for Space Weather Sciences, New Jersey Institute of Technology, 323 Martin Luther King Blvd, Newark, NJ 07102-1982, USA}
\affiliation{Big Bear Solar Observatory, New Jersey Institute of Technology, 40386 North Shore Lane, Big Bear City, CA 92314-9672, USA}
\author[0000-0002-5233-565X]{Haimin Wang}
\affiliation{Institute for Space Weather Sciences, New Jersey Institute of Technology, 323 Martin Luther King Blvd, Newark, NJ 07102-1982, USA}
\affiliation{Big Bear Solar Observatory, New Jersey Institute of Technology, 40386 North Shore Lane, Big Bear City, CA 92314-9672, USA}
\newcommand{\hw}{\color{red}}
\newcommand{\ha}{H$\alpha$}
\accepted{2025-07-22}
\submitjournal{ApJ}
\begin{abstract}
We investigated mini-filament (MF) eruptions near coronal hole (CH) boundaries to explore their role in coronal dynamics and their potential contributions to the solar wind. Using high-resolution H$\alpha$ images from the 1.6m Goode Solar Telescope at Big Bear Solar Observatory and AIA 193 \AA~ EUV data from Solar Dynamic Observatory, we analyzed 28 MFE events over 7.5 hours of observation spanning 5 days. Three largest MF eruptions triggered distinct coronal responses: two consecutive MFEs produced a small-scale eruptive coronal ejection, while the other generated a jet-like brightening. Furthermore, the 25 smaller-scale MFEs were associated with localized brightenings in coronal bright points (CBPs). These findings suggest that MFs play a significant role in transferring mass and magnetic flux to the corona, particularly within CH regions. We found certain trend that the size of MFEs is correlated with the EUV emissions.   In addition, we observed magnetic flux cancellation associated with MFEs. However, except for a few largest MFEs,  quantitative analysis of magnetic field evolution is beyond the capability of the data.  These results underscore the importance of MFEs in the dynamic coupling between the chromosphere and corona, highlighting their potential role in shaping heliospheric structures. 
\end{abstract}

\section{Introduction}

The solar chromosphere is highly dynamic, even in regions of the quiet Sun and around coronal holes (CHs), exhibiting various small-scale activities such as spicules and mini-filament eruptions (MFEs). Mini-filaments (MFs) are smaller analogs of large-scale filaments, typically spanning tens of megameters in length \citep{wang_minifilament_2000}. These dense, thread-like structures are embedded within magnetic flux ropes, and their eruptions release significant amounts of energy and mass. MFEs are considered to be omnipresent and highly recurrent in the chromosphere \citep{sterling_solar_2000, panesar_iris_2018}. Their eruptions can result in a rapid release of plasma and energy into the corona \citep{hong_mini-filament_2016, sterling_another_2022}. For example, \ha~observations from New Vacuum Solar Telescope (NVST) reveal the complex interactions between MFEs and solar jets, highlighting the role of magnetic reconnection in destabilizing the embedded magnetic flux ropes. Statistical studies indicate that MFs typically span less than 25 Mm in length and are generally aligned with the polarity inversion lines (PILs) of small-scale magnetic bipoles. Their lifetimes range from about an hour to nearly a day \citep{wang_minifilament_2000, panesar_magnetic_2017}. Their eruptions are frequently associated with flux cancellation at footpoints, further emphasizing the role of localized magnetic reconnection in triggering these dynamic events \citep{panesar_magnetic_2017, wang_high-resolution_2024}. Recent spectroscopic studies using Fast Imaging Solar Spectrograph (FISS) data have provided detailed diagnostics of MF properties and associated energy release during the eruption \citep{wang_high-resolution_2024}. The observed MF exhibit temperature up to $1.2 \times 10^{4}$ K and rising speed up to $19$ km/s with thermal and kinetic energy contributions each reaching $3.6 \times 10^{24}$ erg and $2.6 \times 10^{24}$ erg, respectively. These high-resolution analyses emphasize the significant role of MFEs in the chromosphere. Statistical studies, such as \citet{huang_statistical_2023}, identified over 27,000 small-scale EUV ejections, including blow-out and standard jets, near CH boundaries, further linking MFEs to solar wind dynamics. Despite their ubiquity, the study of MFs has been constrained by observational limitations resulting from their small size and short lifetime. High-resolution observations, such as from the Goode Solar Telescope at the Big Bear Solar Observatory (BBSO/GST), combined with space observations from  Atmospheric Imaging Assembly aboard the Solar Dynamics Observatory (SDO/AIA), have provided valuable physical insights of them.  Observations have shown that MFEs not only drive localized chromospheric activity but also facilitate mass and energy transport into the corona  \citep{raouafi_magnetic_2023}. Recent advances in spatial and temporal resolution like these are necessary to fully understand the dynamics and impact of these MFEs.

The solar corona exhibits responses to chromospheric activities such as MF eruptions. MFEs are capable of producing various dynamic phenomena, including mini-CMEs, coronal jets, and microflares, which contribute small-scale transients to the overall energy budget of the corona and solar wind acceleration \citep{innes_quiet_2008, raouafi_solar_2016}. These events highlight the critical role of mass and energy transfer from the lower atmosphere, as coronal heating is not solely driven by energy deposition at coronal heights but also relies on persistent upflows across magnetic field configurations from the chromosphere \citep{de_pontieu_observing_2009}. MFEs are frequently accompanied by surges of cool plasma observed in \ha~ and associated with brightening at their footpoints in soft X-ray (SXR) wavelengths which are suggested to be driven by magnetic reconnection between twisted, cool loops and open magnetic field lines, indicates the presence of localized heating and small-scale flares during the eruptions \citep{harrison_x-ray_1988, shibata_observations_1992, alexander_notitle_1999}. A particularly prominent response of the corona to MFEs is their interaction with coronal bright points (CBPs) \citep{hong_coronal_2014, panesar_iris_2018, sterling_coronal-jet-producing_2020}. CBPs, observed as small-scale, bright, blob-like structures in extreme ultraviolet (EUV) and SXR wavelengths, represent magnetic loops connecting opposite polarities in a network field configuration \citep{vaiana_identification_1973, priest_converging_1994}. Confined plasma in CBPs is heated to temperatures of up to $10^6$~K, likely due to magnetic reconnection \citep{priest_converging_1994}, and CBPs are commonly observed in CHs, the quiet Sun, and near active regions, with lifetimes ranging from hours to days \citep{golub_solar_1974, mcintosh_nine_2005}. MFEs occurring within CBPs have been shown to lead to localized heating, plasma flows, and transient brightenings resembling small-scale analogs of two-ribbon flares commonly associated with large-scale filament eruptions \citep{hong_mini-filament_2016}. These events contribute to the mass and energy loading of the coronal atmosphere, enhancing CBP activity and impacting the surrounding magnetic environment. Observations further suggest that MFEs in CBPs are closely tied to small-scale magnetic flux emergence and reconnection events. Such processes destabilize the magnetic field configuration, triggering rapid plasma upflows and localized heating \citep{madjarska_coronal_2019, madjarska_eruptions_2022}. These interactions not only intensify CBPs but also contribute to broader coronal heating through the dissipation of magnetic energy, providing a significant energy source for the maintenance of the solar atmosphere. The detailed dynamics of MFEs and their interaction with CBPs emphasize their importance in linking chromospheric activity to coronal processes and solar wind outflows.

Further studies suggest that MF-driven small-scale ejections contribute to the formation of small-scale magnetic flux ropes (SMFRs), which are often observed as transient structures in the solar wind. The alignment between MFEs and solar wind transients detected by Parker Solar Probe (PSP) provides compelling evidence for their role in shaping heliospheric magnetic structures \citep{raouafi_magnetic_2023, huang_statistical_2023}. Small-scale magnetic flux ropes (SMFRs) observed in the solar wind have garnered significant attention for their potential origins and roles in heliospheric dynamics. SMFRs are coherent, small-scale magnetic structures with durations ranging from tens of seconds to an hour and typical sizes of several hundred kilometers to approximately 0.001 AU \citep{chen_small-scale_2021, chen_small-scale_2022}.  These structures are particularly abundant in the slow solar wind and are thought to play a key role in the transport of mass and energy within the heliosphere. Several studies \citep{fisk_behavior_2001, cravens_plasma_2019} suggest that SMFRs may originate from interchange reconnection near the solar surface. Such reconnection processes are likely associated with small-scale eruptive events, such as MFs near the CH boundaries. During MF eruptions, magnetic flux ropes embedded within these structures can be destabilized and propelled outward along open CH magnetic fields, contributing to the formation of SMFRs in the solar wind. Statistical analyses \citep{huang_statistical_2023} suggest that MF eruptions near CHs can also generate small-scale coronal ejections, which propagate outward along the open magnetic fields in CH, contributing to the magnetic flux ropes in solar wind. They show a close alignment between the occurrence rates of small-scale ejections around CHs and SMFRs detected in heliospace, further supporting the hypothesis that MFs are key sources of these magnetic structures. Understanding the connection between MF eruptions, coronal dynamics, and the solar wind remains critical for unveiling the fundamental processes governing solar-heliospheric interactions. \citet{raouafi_magnetic_2023} further linked MF-driven jetlets and ejections to solar wind transients, emphasizing that their occurrence rates closely match observed switchbacks and small scale magnetic flux ropes in PSP data. Statistical analyses by \citet{huang_statistical_2023} also supported the significant contribution of small-scale eruptions to heliospheric magnetic structures, reinforcing their role in shaping solar wind properties.

In this study, we focus on the observation of MFEs at the boundary regions of CHs and their role in heating overlaying CBPs. We utilized high-resolution imagery from BBSO/GST and SDO/AIA at 193 \AA\ to observe chromospheric activities and their corresponding coronal responses. We conducted detailed studies on three largest events. Additionally, statistical analysis was performed on 28 observed MFEs, integrating information from \citet{huang_statistical_2023} to assess the potential role of these MFEs as SMFRs in the solar wind. We will first present observations and data analysis.  Then both case studies of larger events and statistical analyses of all events are presented.  Finally, we will discuss the possible roles of MFEs in heating solar atmosphere and contributing to transient structures in the solar wind.

\section{Observation and Data Processing}

\subsection{Overview of Observational Data}

For chromospheric activity observations, we used high-resolution H$\alpha$ images from the Visible Imaging Spectrometer (VIS) observed by Goode Solar Telescope (GST) at Big Bear Solar Observatory (BBSO). VIS is equipped to capture detailed chromospheric dynamics, enabling the study of MFs and their eruptions. The VIS instrument provides high temporal resolution ($\sim$24 seconds) and spatial resolution of approximately 0.1\arcsec~per pixel, making it an ideal tool for observing fine-scale chromospheric structures. The data include H$\alpha$ images at the line center and line wings for the observation of the structure and the line-of-sight (LOS) velocity. 

In addition to chromospheric observations, we utilized data from the Near InfraRed Imaging Spectropolarimeter (NIRIS) on BBSO/GST to investigate the underlying photospheric magnetic field evolution. NIRIS provides high-resolution magnetograms, capturing photospheric dynamics critical for understanding the magnetic flux changes driving MF eruptions. We perform area integration of Stokes line profiles along the wavelength, assuming that the field strengths of magnetic features are weak enough (typically $\lesssim$ 1~kG) and the Milne-Eddington atmospheric approximation holds, and retrieve the LOS component of magnetograms. These magnetograms have a spatial resolution of approximately 0.2\arcsec~and a temporal cadence of 30 seconds, enabling the study of rapid magnetic field evolution associated with chromospheric eruptions.

To facilitate future comparisons with small-scale magnetic flux ropes (SMFRs) detected in the solar wind, we selected data from a PSP encounter period during which PSP was within 0.25 AU of the Sun, and the footpoints traced back to the solar surface were positioned on the Earth-facing side of the Sun. Specifically, we analyzed data from PSP Encounter 5, which took place from 2020-06-09 to 2020-06-13. During this period, GST tracked the PSP footpoints on the solar surface, as provided by the Whole Heliosphere and Planetary Interactions (WHPI, \url{https://whpi.hao.ucar.edu/whpi_campaigns.php}) project. These footpoints were calculated to be located at the boundaries of the polar CH region. The GST high-resolution data availability are presented in Table~\ref{data}. VIS provided high-resolution H$\alpha$ images, capturing the line-center and at offsets of $\pm 0.4$ \AA, $\pm 0.8$ \AA, (and on 2020-06-12 and 2020-06-13, at $\pm 1.2$ \AA\ in addition). The cadence for each image set was $\sim$24 seconds, and the near-simultaneous line-center and line-wing images were used to compute pseudo-Dopplergrams, calibrated with standard data from the BAse de données Solaire Sol database (BASS200, \url{https://bass2000.obspm.fr/solar_spect.php}).

\begin{figure}[!ht]
    \centering
    \includegraphics{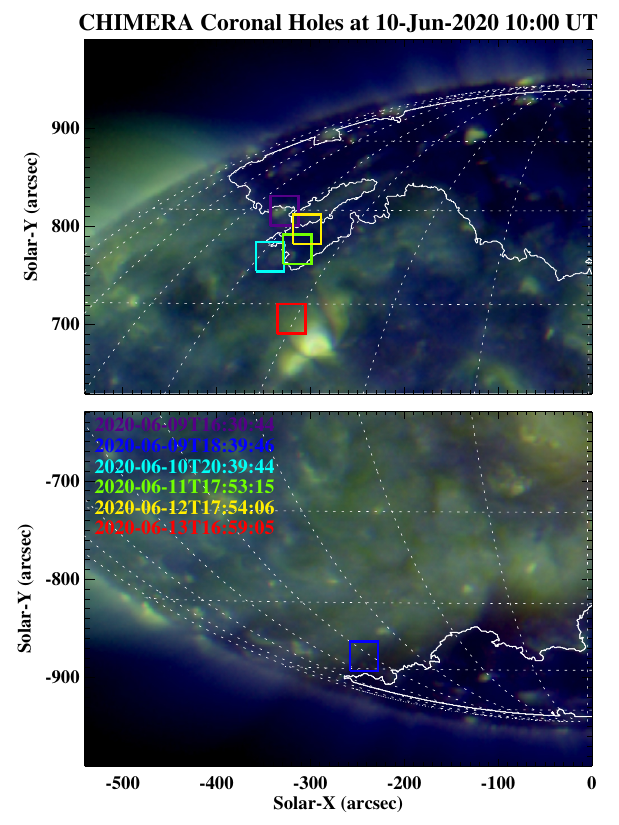}
    \caption{The CH map generated by CHIMERA code. The image is constructed by the emission intensity of three EUV wavelengths (171 \AA, 193\AA~ and 211\AA). The white contours show the boundary lines of each CH. The colored boxes show the re-projected location of our data selected FOV during PSP  Perihelion 5, following predicted PSP footpoints provided by WHPI team.}
    \label{CH}
\end{figure}

Since GST is a ground-based telescope, observations were subject to atmospheric seeing conditions. To maintain consistency, we excluded data affected by poor seeing, retaining only time intervals with stable image quality. In the observation periods listed in Table~\ref{data}, after excluding time intervals with poor seeing conditions, we ultimately obtained 7.5 hours of high-quality data. The GST/VIS field of view (FOV) has a diameter of approximately 62 \arcsec, but due to image quality variations, we used only the central 30 \arcsec\ of each frame for analysis. In Figure\ref{CH}, the GST FOVs selected are denoted using the colored boxes over the CH map generated using CHIMERA code \citep{morgan_multi-scale_2014}. The coordination of each FOV box is rotated to the time of this CH map to show the relative locations to the CHs. To be noted, the adaptive optics (AO) system used by GST requires photospheric structures with sufficient contrast within the FOV to serve as a guide for effective image stabilization. Even during selected observing periods, variations in atmospheric seeing can occasionally degrade image quality, resulting in segments with suboptimal clarity that are unsuitable for detailed analysis. This can cause portions of the data to lack sufficient resolution for observing fine-scale chromospheric structures effectively.

For each observed MF eruption, we analyzed the coronal response using the 193 \AA\ band from the SDO/AIA. The 193 \AA\ channel is sensitive to plasma at temperatures corresponding to those typically found in coronal jets and heated coronal loops, making it well-suited for studying the coronal responses to chromospheric eruptions. By focusing on this wavelength, we could directly observe the energetic transfer from the chromosphere into the corona.

\begin{table}
\begin{center}
\caption{Summary of GST Data Used in This Study}
\label{data}

\begin{tabular}{lcccc}
\\
 \hline\noalign{\smallskip}
Date        &	Observation Period		&	VIS image wavelength	&  NIRIS 	&     FOV center Coordination(\arcsec)	\\
 \hline\noalign{\smallskip}

2020-06-09		&		16:27 -- 17:42		&		0, $\pm 0.4$ Å, $\pm 0.8$ Å		& He 10830 &		[-375\arcsec, 816\arcsec]		\\
 		&		18:15 -- 19:24		&		0, $\pm 0.4$ Å, $\pm 0.8$ Å		&	 He 10830  &	[-273\arcsec, 878\arcsec]		\\
2020-06-10		&		20:00 - 22:43		&		0, $\pm 0.4$ Å, $\pm 0.8$ Å			& Magnetograms &		[-302\arcsec, 771\arcsec]		\\
2020-06-11		&		16:21 -- 21:51		&		0, $\pm 0.4$ Å, $\pm 0.8$ Å			& Magnetograms &		[-183\arcsec, 776\arcsec]		\\
2020-06-12		&		17:45 -- 18:47		&		0, $\pm 0.4$ Å, $\pm 0.8$ Å, $\pm 1.2$ Å		& Magnetograms &		[-84\arcsec, 796\arcsec]		\\
2020-06-13		&		16:53 -- 17:18		&		0, $\pm 0.4$ Å, $\pm 0.8$ Å, $\pm 1.2$ Å		& Magnetograms &		[110\arcsec, 705\arcsec]		\\
 		&		20:06 -- 20:24		&		0, $\pm 0.4$ Å, $\pm 0.8$ Å, $\pm 1.2$ Å		& Magnetograms &		[133\arcsec, 755\arcsec]		\\
 \hline\noalign{\smallskip}\hline
\end{tabular}
\end{center}
\end{table}

\subsection{Identification of Mini-filament Eruptions}

We identified chromospheric eruptions using high-resolution H$\alpha$ images from GST/VIS. As described earlier, each VIS image set includes H$\alpha$ line-center and symmetrically offset line-wing images, enabling the calculation of high-resolution H$\alpha$ pseudo-Doppler maps. To ensure accurate intensity calibration across the multiple H$\alpha$ bands, we processed these images against a relatively quiet subregion of each daily dataset. Assuming the total Doppler shift across these quiet areas averages to zero, we integrated the emission across different wavelengths within these regions.
Using H$\alpha$ reference intensity profiles for line-center and line-wing positions provided by BASS2000, we derived calibration factors specific to each wavelength for each day, following the calculation mentioned in \citet{su_observations_2016}. These calibration factors were then applied to the full set of H$\alpha$ images to standardize intensity values across wavelengths.
For each calibrated image set, we calculated the moments of the line profiles at every pixel. By determining the centroid of each profile, we derived the Doppler velocity at each pixel, effectively mapping the dynamic chromospheric activity with a high level of spatial and temporal precision. Based on the spectral resolution of VIS (0.4 \AA) and typical signal-to-noise ratios in data used in calibration, we estimate a velocity uncertainty of approximately 1.5–2.0 km s$^{-1}$. 

Using the pseudo-Dopplergrams generated from our calibrated data, as shown in Figure~\ref{MF1}(a), we observed a variety of chromospheric features within our selected FOV, including the fine structures of spicules and MFs, with clear LOS Doppler velocity signatures. To identify MFs, we primarily relied on visual inspection of pseudo-Dopplergrams sequences, constructing movies from consecutive images to track the temporal evolution of these features.
For each identified feature, we cross-referenced the pseudo-Dopplergrams with the original VIS images at multiple wavelengths to confirm its structure and dynamics. This verification process ensured that observed features were consistent across different spectral bands and that artifacts or transient seeing effects were excluded from our analysis.

\subsection{Image Registration and Coronal Feature Identification}

The GST data coordinates typically exhibit slight offsets from true solar coordinates, along with small inaccuracies in image rotation angles. For quiet Sun regions, automated image registration often faces difficulties, or even fails entirely. In such cases, we manually identified and corrected both the coordinates and rotation angles of the GST images to align them accurately with full-disk H$\alpha$ images. This manual registration allowed us to map each GST FOV onto the corresponding coronal locations with sufficient precision.

To analyze the coronal response associated with each MFE, we searched for small-scale coronal ejections and brightenings within our FOV during each observation period. In the study by \citet{huang_statistical_2023}, a large number of small-scale coronal ejections were identified, among which two small scale coronal ejection events overlapped with both the observation period and FOV of this study. These two coronal events will be discussed in detail in next section. Due to the variable emission distribution in AIA 193 \AA\ images—particularly with strong differences between CH backgrounds and quiet Sun regions—we employed difference imaging techniques to enhance the visibility of changes in these coronal regions. This approach improved our ability to capture and characterize transient brightenings and structural changes linked to each chromospheric eruption, providing a clearer view of their effects on coronal plasma dynamics.

\section{Results}

Over the five-day observation period, after excluding data affected by seeing disturbances, approximately 7.5 hours of stable high-quality data were analyzed.  Within this period, a total of 28 MFE events were identified. Among these, three MFEs corresponded to two small-scale coronal ejections, which were identified using the algorithm described in \citet{huang_statistical_2023} and enhanced AIA images. We note that the term ‘small-scale coronal ejection’ here follows the definition in \citet{huang_statistical_2023}, referring to localized EUV brightenings and dynamic structures in the low corona that do not have observable white-light CME signatures. These events are distinct from mini-CMEs and should not be interpreted as such due to their substantially small scales.
Following the classification scheme for coronal ejections outlined in \citet{huang_statistical_2023}, one of these small-scale coronal ejections was categorized as an eruptive event per \cite{huang_statistical_2023}. This event corresponded to two consecutive MFEs within a localized region of the FOV. The other ejection was classified as a jet-like event, associated with three spatially adjacent, simultaneously occurring segments. These classifications facilitated an examination of the distinct coronal responses associated with each type of chromospheric eruption, offering valuable insights into the role of these eruptions in transferring mass and energy into the corona.

\subsection{Case Study of Mini-filament Eruptions as  Source of a Coronal Eruptive Event}

\begin{figure}[ht]
    \centering
    \includegraphics[scale=0.55]{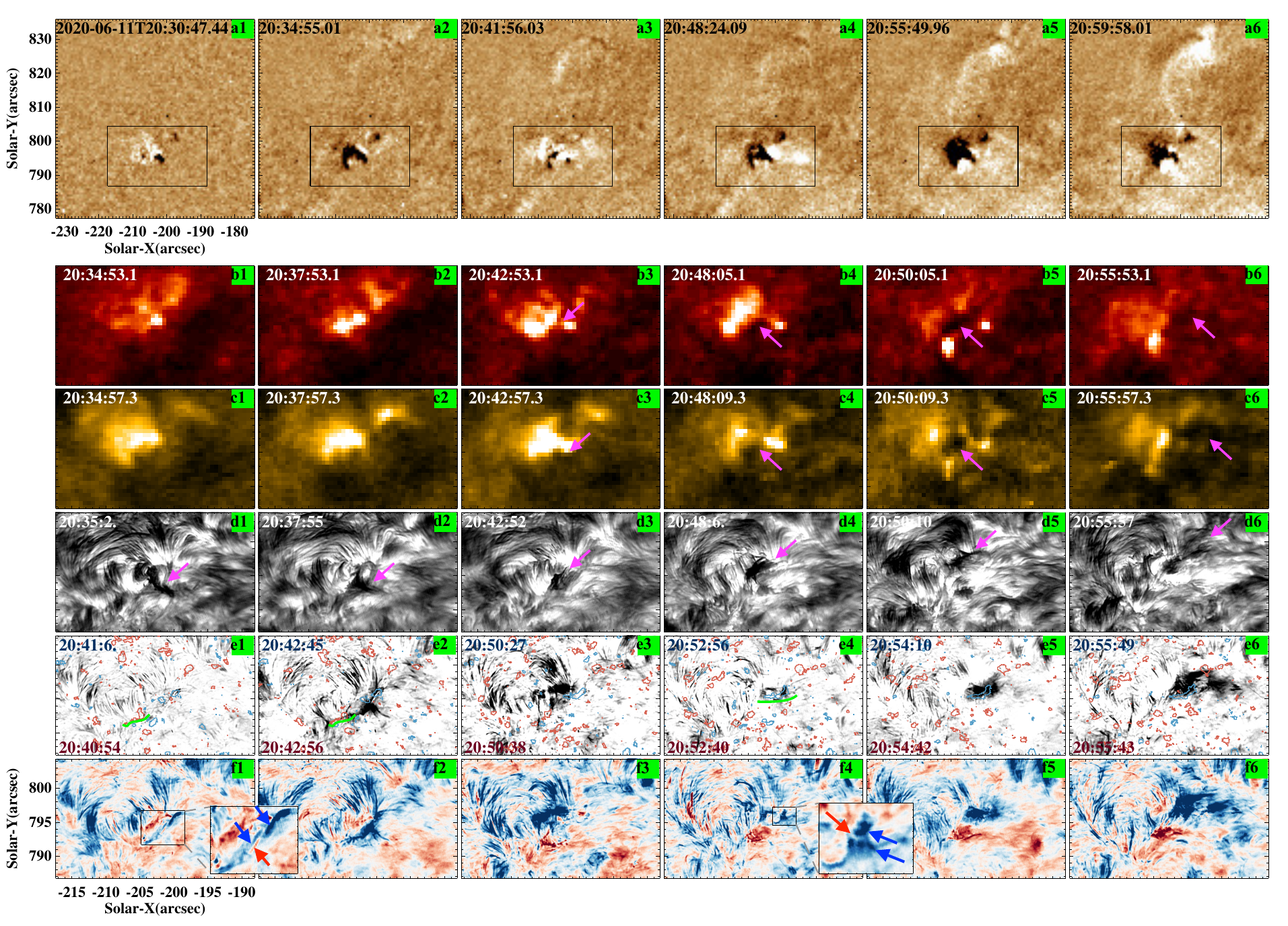}
    \caption{The consecutive two MFEs observed by GST and the corresponding small-scale ejective coronal ejection. (a) Coronal brightening observed by AIA 193 \AA\. The black rectangles indicate the FOV in the lower panels. (b) and (c) AIA 304 \AA\ and 171 \AA\ images. (d)  H$\alpha$ line-center images from VIS. (e) and (f) H$\alpha$ blue-wing images at \ha-0.8\AA\ and Pseudo-Dopplergrams calculated using images at five different wavelengths across the H$\alpha$ line profile. The contours in panels (e) reprensent the LOS magnetic elements over $\pm$20 G. The magneta arrows in the panels (b-d) point out the eruption in the \ha\ line center images. The green lines represent the location where each MF first observed in \ha-0.8 \AA\ images. In the zoom-in panels in (f1) and (f4), The red and blue arrows point out redshifted and blueshifted structure in the MFs during the two eruptions. From left to right, panels 1–3 correspond to the first MFE and panels 4–6 correspond to the second MFE. Panels 1 and 4 show the initial phase of each MF, panels 2 and 5 show their fully developed structures, and panels 3 and 6 capture their eruption phases. An animated version of this figure is available in the online journal. This animation shows difference image of AIA 193 \AA\, images of AIA 304 \AA\ and 171 \AA, H$\alpha$ line-center images and H$\alpha$ blue-wing images at \ha-0.8 \AA\ from VIS, and Pseudo-Dopplergrams calculated using images at five different wavelengths across the H$\alpha$ line profile for the period of 20:30 -- 21:10 UT. (An animation of this figure is available.)}
    \label{MF1}
\end{figure}

During the observation on 2020-06-10, we identified two consecutive MFEs that collectively triggered one small scale blow-out coronal ejection. Figure~\ref{MF1} illustrates the formation and eruption process of these two MFs. 

Each MFE was observed to form at the spicule roots along the granulation boundary, where the initial stages of the eruptions were marked by the gradual ascent of material over approximately 2 min. During this time, each MF grew to an approximate width of 3\arcsec~and length of 5\arcsec, showing an overall upward Doppler shift in velocity. As the eruptions progressed, fine structures within the MFs displayed intertwined redshifted and blueshifted regions, indicative of an untwisting motion within the magnetic flux rope structure. These Doppler-shifted features are highlighted in the pseudo-Dopplergram panels (f1) and (f4) of Figure~\ref{MF1}, where red and blue arrows in the inset zoom-in panels indicate redshifted and blueshifted structures, respectively. We note that in panel (f4), the second MF eruption shows a dominant blueshifted pattern, within which substructures with slightly lower blueshift velocity (pointed out by the red arrow) represent the red shifted part in relative to the blueshifting eruption. Although our temporal resolution limits prevented precise quantification of this untwisting, this phase highlights the gradual buildup of magnetic tension prior to eruption.
The eruptions then entered an acceleration phase, during which the MFs exhibited blueshifted velocities, with plasma rapidly ascending and subsequently dissipating within 2-5 mins as the density decreased, making the features fainter in the images. These two MF eruptions were identified in the H$\alpha$ -0.8 \AA~ images and pseudo-Doppler maps, occurring at 20:40–20:52 UT and 20:52–21:01 UT, respectively, at adjacent but distinct locations, as shown in Figure \ref{MF1}. Figure~\ref{MF1} presents two consecutive MFEs, shown in panels (e1)–(f3) and (e4)–(f6), respectively. The green dashed lines indicate the initial locations where these two consecutive MFEs first appeared in the -0.8 \AA~ blue-wing images. In contrast, during the occurrence of these two consecutive MFEs, the AIA 304 \AA\, 171 \AA\, and H$\alpha$ line-center images in panels (b) to (d) only revealed a single left-to-right eruption event. This discrepancy between the blue-wing and line-center observations suggests that some MFEs observed in H$\alpha$ line-center images may actually be the result of multiple successive eruptions occurring at the MF footpoints.

Simultaneously, faint but consistent brightenings were observed in the corona through AIA 193 \AA~difference imaging, which appeared to correspond to an eruptive coronal ejection  \citep{huang_statistical_2023}. Due to the faintness of the observed features, the coronal ejection associated with these MFEs was less pronounced in EUV observations. This coronal response suggests that the MF eruptions contributed to mass loading and heating in the upper coronal plasma. The low emission intensity observed in this event may be attributed to the relatively low temperature of the MF plasma and the low density of the surrounding coronal environment.

\begin{figure}[ht]
    \centering
    \includegraphics[scale=0.85]{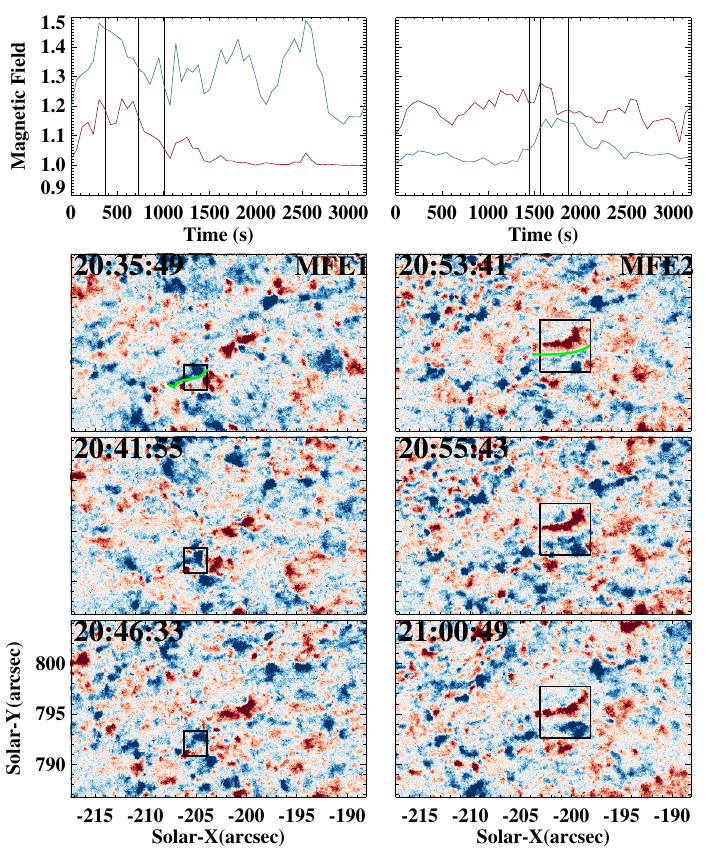}
    \caption{Evolution of the line-of-sight (LOS) magnetic field at the footpoints of the two consecutive mini-filament eruptions (MFEs) presented in Figure~\ref{MF1}. The left column corresponds to the first MFE, while the right column corresponds to the second MFE. The top row shows the time evolution of the integrated positive (red) and negative (blue) magnetic flux within the footpoint region, marked by black rectangular boxes in the lower panels. The second to fourth rows present LOS magnetograms at three key stages: MF formation, eruption, and post-eruption. The black rectangular boxes in these panels indicate the analyzed region. The green dashed lines represent the locations where the MFs emerge, as is in Figure~\ref{MF1}. The three vertical lines in the top row mark the time stamps corresponding to the magnetograms below.}
    \label{MF1mag}
\end{figure}

To further analyze the magnetic field changes during these eruptions, we present Figure~\ref{MF1mag}, which displays the LOS magnetic field evolution at the footpoints of the two MFEs. Utilizing high-resolution and high-sensitivity magnetograms from GST/NIRIS, we detected and tracked the movement and evolution of small-scale magnetic features associated with the MFs. Each of the MFEs start with the emergence of magnetic elements in opposite polarity in the polarity inversion area. Subsequently the cancellation of one polarity  with surrounding opposite polarity flux  coincides with the eruption. Changes in the  magnetic field revealed magnetic emergence and cancellation corresponding to the two MFEs. In Figure~\ref{MF1mag}, the left column corresponds to the first MFE, while the right column corresponds to the second MFE. The first row shows the time evolution of the integrated positive (red) and negative (blue) magnetic flux over $\pm$10 G within the footpoint region (marked by black rectangular boxes in the lower panels). The three vertical lines indicate the time points corresponding to the LOS magnetograms shown below. The second to fourth rows present the LOS magnetograms at three key stages: MF formation, eruption, and post-eruption.
For the first MFE (left column), which occurred in a region dominated by negative polarity, small positive magnetic elements emerged, and the initial location of the MF appeared between the pre-existing negative element and the newly emerged positive element. As the MF erupted, the opposite polarity elements underwent cancellation. This emergence and subsequent cancellation of the positive magnetic flux are reflected in the red curve in the top-left panel.
Similarly, for the second MFE (right column), which occurred in a region dominated by positive polarity, small negative magnetic elements appeared to emerge and subsequently cancel with surrounding positive polarity magnetic fields. The corresponding magnetic flux evolution is shown in the top-right panel, where the blue curve represents the negative flux variations.

Since these eruptions occurred within a coronal hole region, the magnetic flux released during the two MF eruptions may have been ejected into the outer corona, potentially contributing to the solar wind as SMFRs. This scenario underscores the role of MFEs in transferring magnetic flux and energy from the chromosphere into the heliosphere, as observed by PSP. However,  observation is not sufficient to track such small individual coronal ejections to PSP in-situ observations.  Notably, these two events correspond to the “eruptive” category of coronal ejections identified in \citet{huang_statistical_2023}, and we find that only larger MFEs produce such coronal responses. In contrast, smaller MFEs do not generate clear eruptive signatures in AIA 193 \AA\ images. On the other hand, smaller events may cause the similar coronal eruptions, but may be undetected due to sensitivity and resolution of current EUV observations.  

\subsection{ Case study of a Mini-filament Eruption as the Source of a Coronal Jet-like Event}

\begin{figure}
    \centering
    \includegraphics[scale=1]{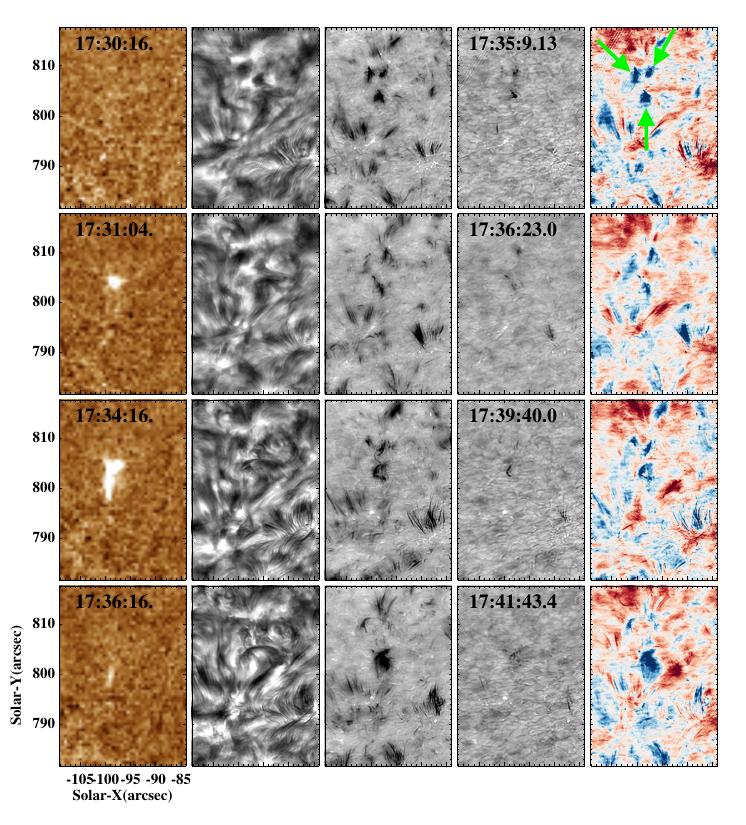}
    \caption{A jet-like coronal ejection and corresponding MFEs. Left column: AIA 193  \AA\ difference images presenting a jet-like Y shape brightening in the corona. Columns 2-4: the VIS data taken at \ha\ line-center, -0.8 \AA\ and -1.2 \AA~. Right column: the pseudo-Doppler maps constructed by images from 7 wavelength. For each row, the \ha\ images are from the same set of data which are taken nearly simultaneously, and the AIA images are the closed frame to the VIS dataset. The green arrows in the top row points to the three MFs at the beginning of the eruption. An animated version of this figure is available in the online journal. This animation shows difference image of AIA 193 \AA, H$\alpha$ line-center images and blue-wing images at \ha-0.8 \AA~ and \ha-1.2 \AA\ from VIS, and Pseudo-Dopplergrams calculated using images at five different wavelengths across the H$\alpha$ line profile for the period of 17:35 - 17:45 UT 
    (An animation of this figure is available.)}
    \label{MF2}
\end{figure}

During observations on 2020-06-12, we detected three adjacent MF components erupting simultaneously from 17:36 to 17:42 UT, as shown in Figure~\ref{MF2}. Unlike the eruptive event discussed in the previous section, where a single MFE triggered a small-scale coronal ejection, this case demonstrates a different type of coronal response: a ``jet-like'' brightening associated with multiple concurrent eruptions of multiple parts of MF. This aligns with the classification in our previous study \citep{huang_statistical_2023}, where we identified two types of small-scale coronal ejections—eruptive events and jet-like events. Here, we present an example of the latter, reinforcing the observational distinction between these two categories. Simultaneous to the MFE, the AIA 193 \AA\ difference images displayed a jet-like brightening in the corona as shown in the left column. The columns 2-4 in Figure~\ref{MF2} show the VIS data taken at \ha\ line-center, -0.8 \AA\ and -1.2 \AA\, while the right column shows the pseudo-Doppler maps constructed by images from 7 wavelengths. 

In the upper row of Figure~\ref{MF2}, the green arrows point to the three MFs at the beginning of the eruption. As is shown, the MFs were spatially arranged in a near-equilateral triangular formation.  In the \ha\ -1.2 \AA\ images, all three MFs displayed strong upward velocities, evidenced by dark absorption features, indicating rapid plasma ascent. However, in the H$\alpha$ line-center images, these MFs did not exhibit well-defined filament structures, which suggests that the eruptions were primarily driven by material at higher Doppler shifts.

A comparison with the co-aligned AIA image reveals that these three MFs were positioned near the footpoints of a reversed-Y structure, specifically at the two legs and the junction of the Y shape. This configuration resembles the morphology of standard coronal jets, where reconnection between closed and open field lines leads to plasma ejection. However, we classify this event as a "jet-like" brightening rather than a full-fledged jet or jetlet because its elongated morphology and coronal response differ from classic jet structures. Instead, the AIA time-series suggests that this brightening is more consistent with the activation of a slender coronal loop system.

Based on H$\alpha$ line-center images and NIRIS magnetograms, we confirmed that these three MFs were independent of each other. However, the simultaneous eruption suggests that a larger-scale structure may have played a role in triggering these events collectively. In contrast to the previous section, where an isolated MFE was sufficient to drive an eruptive ejection, this case demonstrates that jet-like brightenings can emerge when multiple MFEs erupt in close spatial and temporal proximity. This suggests that a shared magnetic environment, possibly a pre-existing sheared field or a local flux rope system, may facilitate the nearly synchronous activation of these MFEs.

The observational characteristics of this event align with our previous classification of jet-like brightenings, reinforcing that these coronal responses are often associated with multiple concurrent MFEs rather than singular large-scale filament eruptions. Importantly, only larger MFEs appear to produce these jet-like brightenings, while smaller MFEs generally exhibit no clear coronal response. This distinction further supports the classification in \citet{huang_statistical_2023}, where only sufficiently energetic MFEs were found to drive observable coronal dynamics.  The proximity of the MFs to the jet-like features footpoints suggests a complex interaction between chromospheric and coronal dynamics, where local reconnection events driven by multiple MFEs may contribute to energy and mass transfer into the corona.

Our observed jet-like events, as captured predominantly in AIA 193 \AA\ images and complemented by high-resolution \ha\ observations, share notable morphological similarities with the small-scale jet-like phenomena commonly referred to as ``campfires" observed by EUI on Solar Orbiter \citep{berghmans_extreme-uv_2021, panesar_magnetic_2021, panesar_solar_2023}. These campfires are characterized by their transient brightenings and dynamic small-scale eruptions in the quiet solar corona, typically located at magnetic network boundaries. 
Our observations align well with these previous studies, specifically in the presence of MFs and associated magnetic flux dynamics. The fine-scale eruptive features and coronal brightenings observed in our study likely result from similar underlying mechanisms involving magnetic reconnection triggered by flux cancellation. Additionally, our results complement the findings of \citet{tiwari_fine-scale_2019}, who reported surge/jet-like events exhibiting untwisting motions within small-scale active regions observed by Hi-C 2.1. These Hi-C observations provided direct evidence of intertwined redshifted and blueshifted Doppler motions, consistent with our observation of untwisting flux ropes.

Therefore, the jet-like events discussed in our work contribute to the growing evidence suggesting that MFEs and associated small-scale coronal brightenings represent a ubiquitous physical mechanism operating at various scales across the solar atmosphere. The detailed high-resolution observations presented in our study, particularly the chromospheric structure in GST/VIS \ha\ images, further highlight the complex magnetic interactions and localized heating processes at play in these small-scale solar eruptive events.

\subsection{Statistical analysis of Smaller MFEs and Corresponding Coronal Bright Points}

\begin{figure}
    \centering
    \includegraphics[scale=0.6]{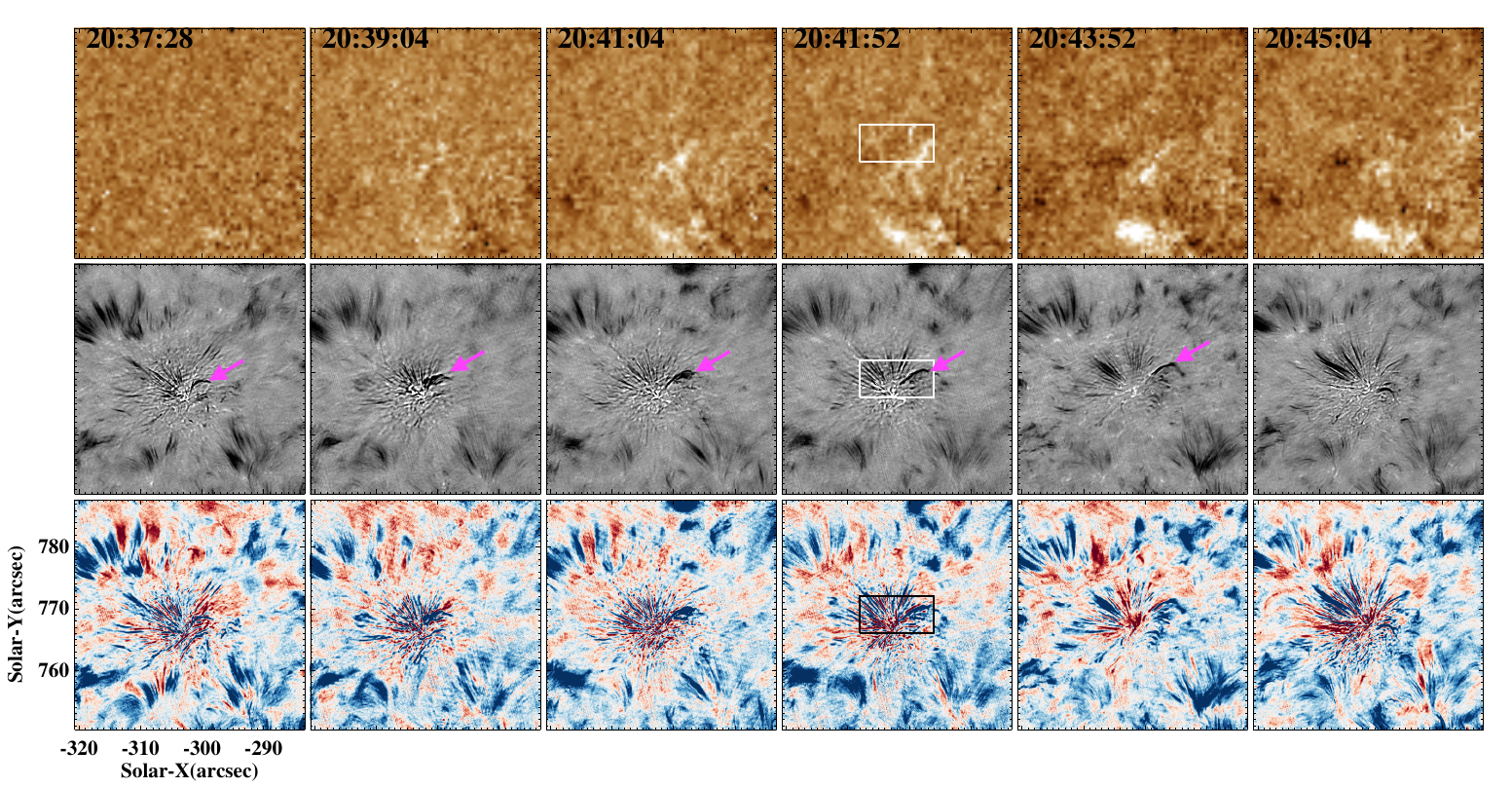}
    \caption{Formation, growth, and eruption of a mini-filament observed using GST/VIS H$\alpha$ images. The top panels:  AIA 193 \AA\ difference images corresponding to the same FOV of GST/VIS. Lower two panels: H$\alpha$ images at \ha $-0.8$ \AA ~and corresponding pseudo-Doppler maps. The erupting MF is pointed out by the magneta arrows in the \ha $-0.8$ \AA~images.}
    \label{micro}
\end{figure}

In addition to the two sets of larger MFs studied above, which produced clear small-scale coronal ejections, we observed 25 additional MFEs over the 6.2 hours of data. Figure~\ref{micro} illustrates one such MFE, capturing its formation, eruption, and eventual dissipation. The erupting MF is marked by magenta arrows in the H$\alpha$ $-$0.8~\AA\ panels. These MFs were observed to form along the roots of spicules near the granulation boundary, oriented perpendicularly to the spicule direction. The footpoints of the MFs were similarly located along the granulation boundary. Following their formation, these structures typically underwent a sudden upward motion and faded from view within 2--5 minutes. Due to limitations in temporal resolution, we were unable to fully resolve the rapid acceleration phase during their ascent.

As the MFs disappeared, broader (approximately 2--3\arcsec\ wide) upward plasma flows were observed along the spicule direction, occasionally exhibiting twisting structures. The top panels of Figure \ref{micro} show AIA 193 \AA~difference images corresponding to the FOV. Unlike the larger MFEs discussed in previous sections, these smaller-scale eruptions did not produce obvious coronal ejections. Instead, their impact on the corona was subtle and often difficult to identify in the highly dynamic EUV images. However, by precisely determining the time and location of these small MFEs, we were able to establish a clear correspondence between their occurrence and localized brightenings as the overlying transient CBPs. 

\begin{figure}
    \centering
    \includegraphics[scale=1.5]{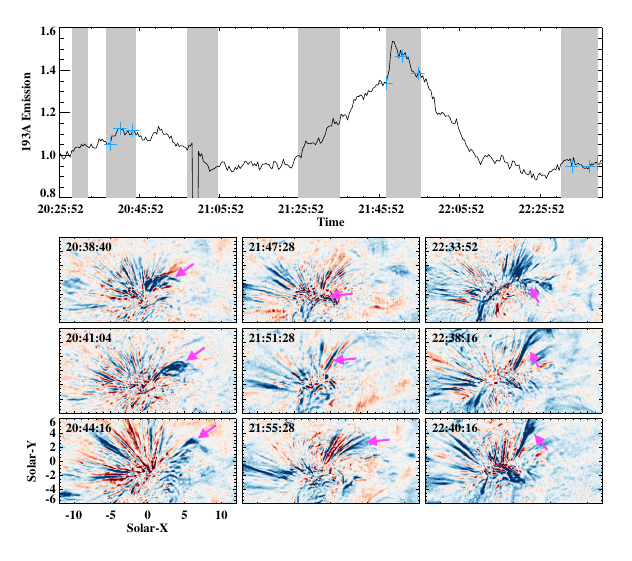}
    \caption{(Top panel) Integrated AIA 193 \AA\ emission profile in the region marked by the white box in Figure~\ref{micro}, with shaded regions corresponding to the timing of six mini-filament eruptions (MFEs). (Bottom panels) Three representative MFE events showing H$\alpha$ pseudo-Doppler map images for different eruption phases: initial acceleration, peak eruption, and post-eruption, and the magenta arrows highlight the MFs during each phases. The blue crosses in the top panel indicate the specific times of the images shown in the bottom panels.}
    \label{loops}
\end{figure}

Figure~\ref{loops} displays the relationship between six such MFEs identified using GST/VIS from 20:25 UT to 22:45 UT on 2020-06-10 and their corresponding EUV emission. The top panel of Figure~\ref{loops} shows the integrated AIA 193 \AA\ emission profile in the region marked by the white box in Figure~\ref{micro}, with shaded areas indicating the time intervals associated with these six MFEs. Images from three representative MFEs are shown in the lower panels of Figure~\ref{loops}. The first through third columns correspond to these three MFEs, and each column's first through third rows show images taken during the initial acceleration phase, the peak eruption phase, and the post-eruption phase, respectively. The specific times corresponding to each image are marked in the top panel by blue crosses. The timing of each MFE eruption closely matched the brightening phase in the CBP system. The dimming phase of the CBPs generally began when the H$\alpha$ plasma flow faded, unless subsequent MFEs occurred.

In all 28 identified MFE events, the eruptions were accompanied by enhanced integrated AIA 193 \AA~emission. However, for the 25 smaller MFEs, the contrast of these brightenings was often low, making them difficult to detect in raw EUV images. The highly dynamic background variability in AIA 193 \AA\ images, especially in coronal holes, further complicates the identification of such small-scale brightenings. By constraining the spatial and temporal occurrence of these MFEs, we were able to more reliably associate them with subtle CBP brightenings, providing evidence that even small MFEs contribute to localized coronal heating.

To further quantify this relationship, we analyzed the length of each MFE and its corresponding EUV brightening. Using VIS H$\alpha$ blue-wing images, we identified the filament length at the starting moment of its eruption. For each MFE event, we computed the integrated EUV contrast in AIA 193 \AA\ as a measure of its coronal response. This was done by first calculating the contrast enhancement within the brightening region and then taking the square root of this value. The resulting quantity was then integrated over the duration of the MFE to obtain the total EUV response. The integrated EUV contrast has a unit of arcsec$\cdot$s and provides a refined characterization of the EUV brightening associated with each MFE.

\begin{figure}[ht]
    \centering
    \includegraphics[scale=1]{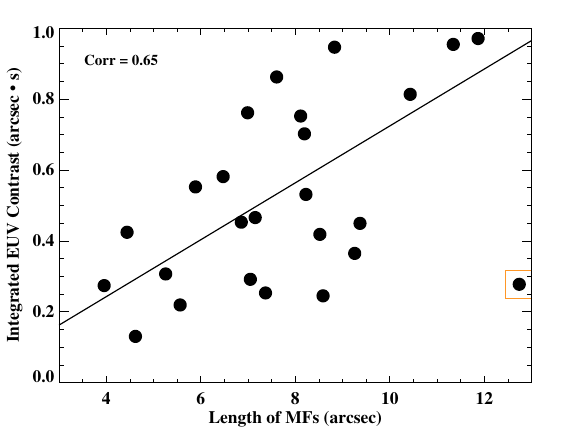}
    \caption{Scatter plot showing the relationship between the length of MFEs and the corresponding integrated EUV contrast in overlying coronal bright points (CBPs). The length measurements are based on \ha $-0.8$ \AA~images from VIS. The solid line shows the linear relationship between them. A moderate positive correlation is observed (correlation coefficient = 0.65).  An event marked by the orange box is a clear outliner, we exclude it in the fitting.}
    \label{length}
\end{figure}

Figure~\ref{length} presents a scatter plot of MFE length versus EUV brightening. The length measurements are based on \ha $-0.8$ \AA~images from VIS. After excluding an obvious outliner event, the results show a moderate  positive correlation (correlation coefficient = 0.65), suggesting that larger MFEs tend to be associated with stronger CBP brightenings. Additionally, we found that the majority of the smaller MFEs observed in our dataset had lengths around 8\arcsec, highlighting a characteristic scale for these features, observed with such high resolution of GST.

\section{Summary and Discussions}

In this study, we analyzed mini-filament (MF) eruptions near coronal hole (CH) boundaries using high-resolution H$\alpha$ images from BBSO/GST VIS and AIA 193 \AA~data. Our observations focused on the interaction between chromospheric eruptions and coronal dynamics, particularly their contributions to coronal heating, mass loading, and small-scale magnetic flux transport.

Over the five-day observation period, we obtained approximately 7.5 hours of usable data within a 30\arcsec~$\times$30\arcsec~FOV. We identified 28 MFE events, including three larger MF eruptions that exhibited distinct coronal responses. Two of these MFs erupted consecutively triggered a small-scale eruptive coronal ejection, while the other produced a jet-like coronal brightening. These findings underscore the role of MFs in driving small-scale coronal ejections and transferring mass and magnetic flux into the corona. Their occurrence within CH boundaries highlights their potential contribution to solar wind structures, such as SMFRs observed in-situ by PSP.

The remaining 25 MFEs were associated with localized brightenings in overlying CBPs without producing observable coronal ejections. These brightenings suggest that MFs may act as energy sources for localized coronal heating, contributing to the dynamic evolution of CBPs. Unlike larger MFEs that generate visible eruptions, these smaller MFEs primarily enhance coronal activity through localized heating, which is often difficult to detect against the dynamic EUV background. By constraining their spatial and temporal properties, we demonstrated a statistical correlation between smaller MFEs and CBP brightenings.

Our investigation of the temporal relationship between chromospheric MFEs and CBP brightenings revealed a consistent pattern: the peak brightening phase in CBPs often coincided with MFE eruptions, while the dimming phase typically followed the dissipation of H$\alpha$ plasma flows. These observations support the hypothesis that MFEs contribute to localized mass and energy transfer processes within the lower corona, although their effects are more subtle compared to larger-scale events.

We calculated the occurrence rate of MFEs based on the 7.5 hours of effective observation time and the selected FOV. The derived occurrence rate per unit time per unit area is approximately:

\[
\text{Occurrence Rate} = \frac{28}{7.5 \times 484 \, \text{Mm}^2} \approx 7.7 \times 10^{-3} \, \text{h}^{-1}\text{Mm}^{-2}
\]

Assuming a uniform distribution across the solar surface (with an approximate area of $6.1 \times 10^6$ Mm$^2$), this translates to a global occurrence rate of$\sim 1.1 \times 10^4$ MFEs per day.

This estimate can be compared to other coronal dynamic events:
\begin{table}[ht]
    \centering
    \caption{Occurrence rates and energy estimates of various solar transients.}
    \begin{tabular}{lcc}
    \hline\hline
        Transients & Occurrence Rate (day$^{-1}$) & Energy/Event (erg)\\
        \hline
        X-ray Jets\footnote{\cite{shibata_observations_1992, savcheva_study_2007, sako_statistical_2013, moore_magnetic_2015, raouafi_solar_2016}} & $\sim10^{2-3}$ & $10^{25} - 10^{27}$\\
        Coronal ejections in CH\footnote{\cite{huang_statistical_2023}} & $\sim2300$ & -- \\
         Jetlets\footnote{\cite{panesar_iris_2018, kumar_quasi-periodic_2022, raouafi_magnetic_2023}}& $\sim2500-10^{5}$ & $\sim10^{24}$\\
        MFEs\footnote{\cite{wang_high-resolution_2024}} & $\sim10^4$ & $\sim10^{25}$ \\
       Spicules \footnote{\cite{sterling_solar_2000, de_pontieu_origins_2011, goodman_acceleration_2012}} & $\sim~10^{8}$ & $10^{22}-10^{25}$ \\
        \hline
    \end{tabular}
    \label{ejections}
\end{table}

The comparisons suggest that MFEs may contribute to a significant fraction of coronal transients in combination with X-ray jets , jetlets and small-scale coronal ejections.
They are over an order of magnitudes more numerous than coronal X-ray jets, and nearly an order of magnitude more than the rates of coronal EUV ejections. It is worth noting that the jetlet occurrence rate in the literature spans more than an order of magnitude, which is likely arise from differences in event–selection criteria (e.g., whether the smallest “campfire–like” brightenings are included), observing passbands and cadences, and the thresholds used for detection. 
On the other hand, the birthrate of spicules are four orders of magnitudes higher. If 10\% of spicules are type II spicules, they are still 1,000 more numerous than MFEs.  This conclusion is based on current resolution of GST/VIS. The short lifetimes and small spatial scales of some MFEs highlight the limitations of current observational capabilities. Improved temporal and spatial resolution is necessary to fully resolve these events and uncover their fine-scale dynamics.

To estimate the energetic impact of these MFEs, we compare our sample with the MFE analyzed by \citet{wang_high-resolution_2024}. In their work, a single MFE was analyzed using high-resolution GST/FISS observations, providing direct measurements of thermal and kinetic energies, specifically, a thermal energy of $3.6 \times 10^{24}$ erg and kinetic energy of $2.6 \times 10^{24}$ erg. They reported MF sizes of $\sim$5 arcsec (length) $\times$ 2 arcsec (width). Our observed MFE sizes range from 4–13 arcsec, with a peak around 8 arcsec, slightly larger than the event analyzed by  \citet{wang_high-resolution_2024}.  Assuming similar thickness (2 arcsec), we approximate the relative scaling of energy and mass using the reported values from \citet{wang_high-resolution_2024}:

The thermal energy scaling based on filament area for the MFEs in our event list is
$E_{\text {thermal}} \sim (8/5) \times 3.6 \times 10^{24} \approx 5.8\times 10^{24} \mathrm{erg}$.
Similarly, kinetic energy per event in the list $E_{\text {kinetic}} \sim (8/5) \times 2.6 \times 10^{24} \approx 4.2\times 10^{24} \mathrm{erg}$. These result in the approximate total delivering energy of $10^{25} \mathrm{erg}$ per event.
With a global occurrence rate of $\sim 1.1 \times 10^4$ MFEs per day, this results in an energy input of
$$
\sim 10^{29} \mathrm{erg} / \text { day }
$$
which is comparable to the estimated total energy requirement for CBPs \citep{madjarska_coronal_2019}.

If we assume an average mass per event of $\sim 10^{12}$ g, scaling based on \citet{wang_high-resolution_2024}, the total mass input from smaller MFEs is estimated to be
$$
\sim 10^{16} \mathrm{~g} / \mathrm{day}
$$

The above calculations indicate that the energy output for MFEs is much smaller than the $10^{32}$ ergs/day of energy requirement for coronal heating. Spicules may provide comparable amount of energy.  On the other hand,  the mass supply of MFEs is much larger than the mass loss of solar wind in the order of $10^{14}$ g/day. However, not all MFEs fully erupt--some undergo only partial eruptions, where only a fraction of their mass is expelled into the corona, while the rest remains confined near the filament channel or falls back due to gravity. Furthermore, only a small fraction of MFEs occurring in CHs or at their boundaries have potential to be funneled into open magnetic fields to contribute to solar wind.    

Our previous study suggested that some MFEs may contribute to SMFRs observed by PSP. However, not all MFEs necessarily evolve into SMFRs. Given that smaller MFEs primarily contribute to localized coronal heating rather than large-scale eruptions, their direct role in SMFR formation may be limited. We suggest that only the larger MFEs (i.e., those producing visible ejections) have a significant chance of contributing to solar wind structures. Further statistical comparisons between PSP SMFR data and our newly derived global MFE rates will be necessary to refine this connection.

Future studies should focus on high-cadence, multi-wavelength observations to better understand the detailed energy conversion processes associated with MFEs. Combining ground-based observations (BBSO/GST, DKIST) with space-based EUV and in-situ measurements (SDO/AIA, Solar Orbiter, PSP) will be essential for determining the full contribution of MFEs to coronal heating and solar wind mass transport.

\section*{Acknowledgements}
We gratefully acknowledge the use of data from the Goode Solar Telescope (GST) of the Big Bear Solar Observatory (BBSO). BBSO operation is supported by  NSF grants AGS-2309939 and New Jersey Institute of Technology. GST operation is partly supported by the Korea Astronomy and Space Science Institute and the Seoul National University. This work was supported by NSF grants, AGS-2114201, AGS-2229064 and AGS-2309939, and NASA grants, 80NSSC19K0257, 80NSSC20K0025, 80NSSC20K1282, 80NSSC241914 and 80NSSC24K0258. The authors benefited significantly from the discussion with Dr. Jeongwoo Lee.


\end{document}